\documentclass[%
 aip,
 amsmath,amssymb,
 reprint,
]{revtex4-2}

\usepackage{graphicx}
\usepackage{dcolumn}
\usepackage{bm}
\usepackage[dvipsnames]{xcolor}
\usepackage[utf8]{inputenc}
\usepackage[T1]{fontenc}
\usepackage{mathptmx}
\usepackage{upgreek}
\usepackage{etoolbox}
\usepackage{multirow}
\usepackage{svg}
\usepackage[english]{babel}
\usepackage{soul}
\usepackage[utf8]{inputenc} 

\makeatletter
\def\@email#1#2{%
 \endgroup
 \patchcmd{\titleblock@produce}
  {\frontmatter@RRAPformat}
  {\frontmatter@RRAPformat{\produce@RRAP{*#1\href{mailto:#2}{#2}}}\frontmatter@RRAPformat}
  {}{}
}%
\makeatother

\usepackage{xr}
\makeatletter
\newcommand*{\addFileDependency}[1]{
  \typeout{(#1)}
  \@addtofilelist{#1}
  \IfFileExists{#1}{}{\typeout{No file #1.}}
}
\makeatother

\newcommand*{\myexternaldocument}[1]{
    \externaldocument{#1}
    \addFileDependency{#1.tex}
    \addFileDependency{#1.aux}
}

\myexternaldocument{SupplyInfo}

\usepackage{hyperref}

\begin{document}

\preprint{AIP/123-QED}

\newcommand{\zy}[1]{{\color[rgb]{.6,.1,.7}{#1}}}
\newcommand{\lv}[1]{{\color{green}{#1}}}
\newcommand{\AM}[1]{{\color[rgb]{1,.0,.0}{#1}}}
\newcommand{\zyc}[1]{{\color[rgb]{.1,.4,.7}{[Joey: {\it #1}\,]}}}
\newcommand{\zyx}[1]{{\color[rgb]{.6,.2,.5}{\st{#1}}}}

\title{Miniaturized and robust tunable monochromatic magneto-optical\,platform for pulsed magnetic fields}
\author{Lev Bergsma}
\affiliation{Institute for Solid State Physics, The University of Tokyo, Kashiwa, Chiba, 277-8581, Japan}
\affiliation{Laboratory for Quantum Magnetism, École Polytechnique Fédérale de Lausanne, CH-1015 Lausanne, Switzerland.}

 \author{Zhuo Yang}%
 \email{zhuo.yang@issp.u-tokyo.ac.jp}
 \affiliation{Institute for Solid State Physics, The University of Tokyo, Kashiwa, Chiba, 277-8581, Japan}

  \author{Bei Sun}%
 \affiliation{Institute for Solid State Physics, The University of Tokyo, Kashiwa, Chiba, 277-8581, Japan}
 
 \author{Yasuhiro H. Matsuda}
 \affiliation{Institute for Solid State Physics, The University of Tokyo, Kashiwa, Chiba, 277-8581, Japan}
 
 \author{Koichi Kindo}
 \affiliation{Institute for Solid State Physics, The University of Tokyo, Kashiwa, Chiba, 277-8581, Japan}
 
\author{Hiroshi Kageyama}
\affiliation{Department of Energy and Hydrocarbon Chemistry, Graduate School of Engineering, Kyoto University, Kyoto 615-8510, Japan.}

\author{Hiroaki Ueda}
\affiliation{Co-Creation Institute for Advanced Materials, Shimane University, Shimane 690-8504, Japan.}

\author{Henrik M.Ronnow}
\affiliation{Laboratory for Quantum Magnetism, École Polytechnique Fédérale de Lausanne, CH-1015 Lausanne, Switzerland.}

\author{Atsuhiko Miyata}
\email{a-miyata@issp.u-tokyo.ac.jp}
\affiliation{Institute for Solid State Physics, The University of Tokyo, Kashiwa, Chiba, 277-8581, Japan}
%


\date{\today}

\begin{abstract}
Tunable monochromatic  magneto-transmission is one of the most established magneto-optical techniques, particularly well suited for pulsed magnetic fields. It employs fixed-wavelength monochromatic light as the probe, while the magnetic field is swept to bring the sample into resonance with the photon energy. The key component of this setup is a tunable laser system, typically consisting of a Ti:sapphire laser coupled with an optical parametric oscillator. However, such laser systems are often bulky, expensive, and inherently unstable, which significantly limits their widespread application in magneto-optical laboratories. In this work, we develop a high-accuracy, cost-effective, and compact tunable monochromatic magneto-transmission system based on a combination of a laser-driven white light source and a mini monochromator, and demonstrate its feasibility and performance in a millisecond range pulsed magnetic field condition. To verify the accuracy of this new and simplified setup, we performed Faraday rotation measurements on the geometrically frustrated spin system $\rm CdCr_2O_4$, as well as magneto-transmission experiments on the Shastry–Sutherland lattice antiferromagnet SrCu$_2$(BO$_3$)$_2$. These results show excellent agreement with previous reports, confirming the reliability and precision of the new setup.

\end{abstract}

\maketitle

\section{\label{sec:level1}Introduction}

Magneto-optical experiments have long been shown to be a powerful tool for probing the electronic band structure\,\cite{nicholas2013ultrahigh} and excitonic properties\,\cite{Yang2017ACSe,miyata2015direct,yang2017unraveling} of materials. By utilizing linearly or circularly polarized light, it also enables access to the magnetic properties known as Faraday and Kerr rotation\cite{Crassee_2010,Higo_2018} and the magnetic circular dichroism, and to the symmetry of magnetic and crystal structures through angle-resolved polarization experiments  \cite{fedchenko2023observationtimereversalsymmetrybreaking}. Recently, the emergence of atomically thin layered materials, where macroscopic experiments are quite challenging, requires magneto-optical experiments to study exotic properties originating from a single-layer sample (e.g., CrI$_3$, NiI$_2$, CrSBr, FePS$_3$) \cite{Song_2022,Jiang_2023,Huang_2017,Tabataba_Vakili_2024,zhou2024giant}. The emergent phenomena in these atomically thin materials—such as field-induced phase transitions and Landau-level spectroscopy—typically occur under high magnetic fields, where pulsed magnetic fields become essential. However, the use of pulsed magnetic fields, with durations typically on the order of milliseconds, imposes strict requirements on the temporal resolution of optical measurements.

\begin{figure*}
  \includegraphics[scale=0.75]{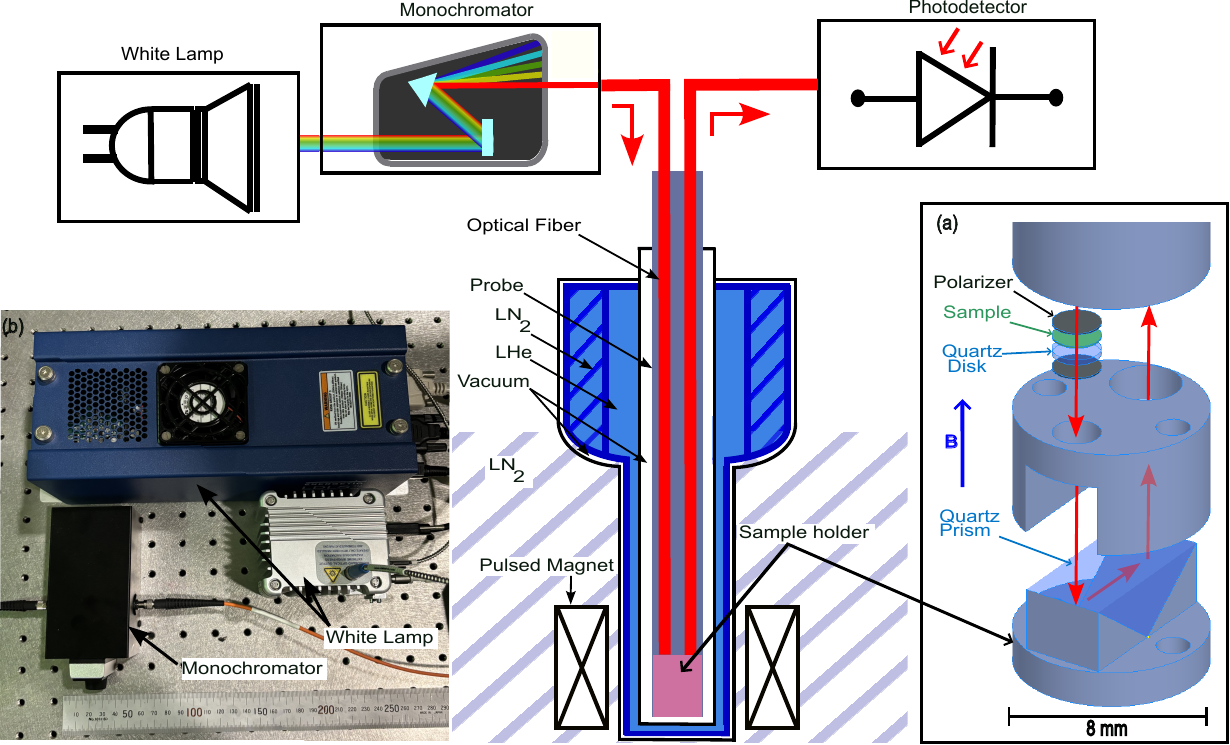}
  \caption{
  Schematic view of the tunable monochromatic magneto-transmission measurement setup. A broadband laser-driven white light source is coupled to a mini-monochromator (top), enabling wavelength-selective illumination. The monochromatic light is delivered via an optical fiber into a probe inserted into a cryostat, which cools the sample to cryogenic temperatures. A pulsed magnet positioned outside the cryostat generates magnetic fields up to 60\,T. The transmitted light is collected by a second optical fiber and directed to a photodetector. Panel (a) shows the sample holder in the Faraday configuration: the incident light passes through a polarizer, the sample, and a quartz disk substrate, then is reflected back by a quartz prism along the same path toward the detector. Panel (b) presents a photograph of the highly compact light source unit, combining the white light source and monochromator within a total length of less than 30\,cm.
  }
\label{fig:EXP} 
\end{figure*}

Among the various optical techniques available for pulsed magnetic field, magneto-transmission is the most widely employed method for probing the optical properties of materials in the visible to near-infrared region. This technique is generally classified into two categories: spectrally-resolved and tunable monochromatic magneto-transmission techniques. The spectrally-resolved method employs broadband light as a light source, and the transmitted light through the sample is dispersed and recorded by a spectrometer. Each transmission spectrum is obtained in a short time scale compared to the pulse field duration, in which case, the spectrum can be acquired at essentially a constant magnetic field. The sampling rate of this technique is around 0.5\,-\,1\,kHz. In contrast, the tunable monochromatic method employs a narrow-band, wavelength-tunable light source (e.g., a tunable laser), and the transmitted light intensity at a fixed wavelength is recorded as a function of magnetic field using a high-speed photodetector with the sampling rate in the order of 100\,kHz. This approach is particularly advantageous under pulsed magnetic field conditions. The fast sweeping rate of the pulsed magnetic field enables efficient tuning across sample resonances, allowing the photon energy to be dynamically matched to field-induced transitions/absorptions. Meanwhile, the high temporal resolution of the photodetector  ensures accurate acquisition of the transmitted signal without loss during the critical time window, circumventing the readout delays associated with spectrally-resolved methods. 
    
The traditional implementation of tunable monochromatic magneto-transmission relies on a wavelength-tunable light source, generally realized through a combination of a Ti:Sapphire laser and an optical parametric oscillator (OPO), providing a high-power and broadly tunable monochromatic beam. While this technique is highly effective—particularly in destructive pulsed magnetic field experiments where the high power is essential due to the extremely short pulse duration (less than 10\,$\upmu$s)\,\cite{YAO2001407,miyata2015direct,Galkowski_2016,Yang2017ACSe} — it suffers from several drawbacks. In addition to the considerable cost and bulky size of such systems, the Ti:Sapphire laser + OPO setup is often prone to operational instability, including sensitivity to environmental temperature and alignment drift. In contrast, for non-destructive magnets, relatively low laser power is enough as their pulse duration is 4 orders higher than that of the destructive pulsed magnet system. For these reasons, we  use a broadband white lamp coupled to a mini monochromator to produce a wavelength-tunable monochromatic light suitable for such conditions. This alternative approach enables tunable monochromatic magneto-transmission measurements with significantly improved cost-efficiency, space savings, and operational stability.

To demonstrate the performance and reliability of this new setup, we measured the magnetization of a geometrically frustrated spin system $\rm CdCr_2O_4$ through the Faraday rotation techniques - one of the most commonly used tunable monochromatic magneto-transmission methods in pulse magnetic field. The measured Faraday rotation angles clearly exhibit a robust magnetization plateau, with a sharp phase transition near 28\,T, in excellent agreement with previously reported results\,\cite{PhysRevB.77.212408}. In addition, we investigated the band-edge behavior of the Shastry-Sutherland lattice antiferromagnet SrCu$\rm_2$(BO$\rm_3$)$\rm_2$ (SCBO) using monochromatic magneto-transmission measurements near 735\,nm. The transmission data reveal a series of sharp features corresponding to field-induced phase transitions associated with magnetization plateaus\,\cite{Matsuda_2013}. The reproducibility, spectral resolution, and signal clarity observed in both measurements strongly validate the accuracy and effectiveness of our experimental setup.

\section{Experimental Method}

 A  general schematic view of the magneto-optical setup is shown in Fig.~\ref{fig:EXP}. A non-destructive pulse magnet was used to generate magnetic field up to 52\,T with a pulse duration of 36\,ms. The sample was placed in an optical probe equipped with two 800\,$\upmu$m diameter multi-mode optical fibers and a prism. One fiber was used to guide the incident light to the sample. The transmitted light from the sample was reflected by the prism to the second fiber that was connected to the detector. The optical path at the bottom of the probe is shown in Fig.~\ref{fig:EXP}(a). The temperature of the sample was monitored by a calibrated Cernox thin-film resistive thermometer mounted at a distance of 1\,mm from the sample; the value of the magnetic field during the measurement was recorded by a calibrated pick-up coil wound close to the sample. 

 \begin{figure} 
\includegraphics[scale=0.32]{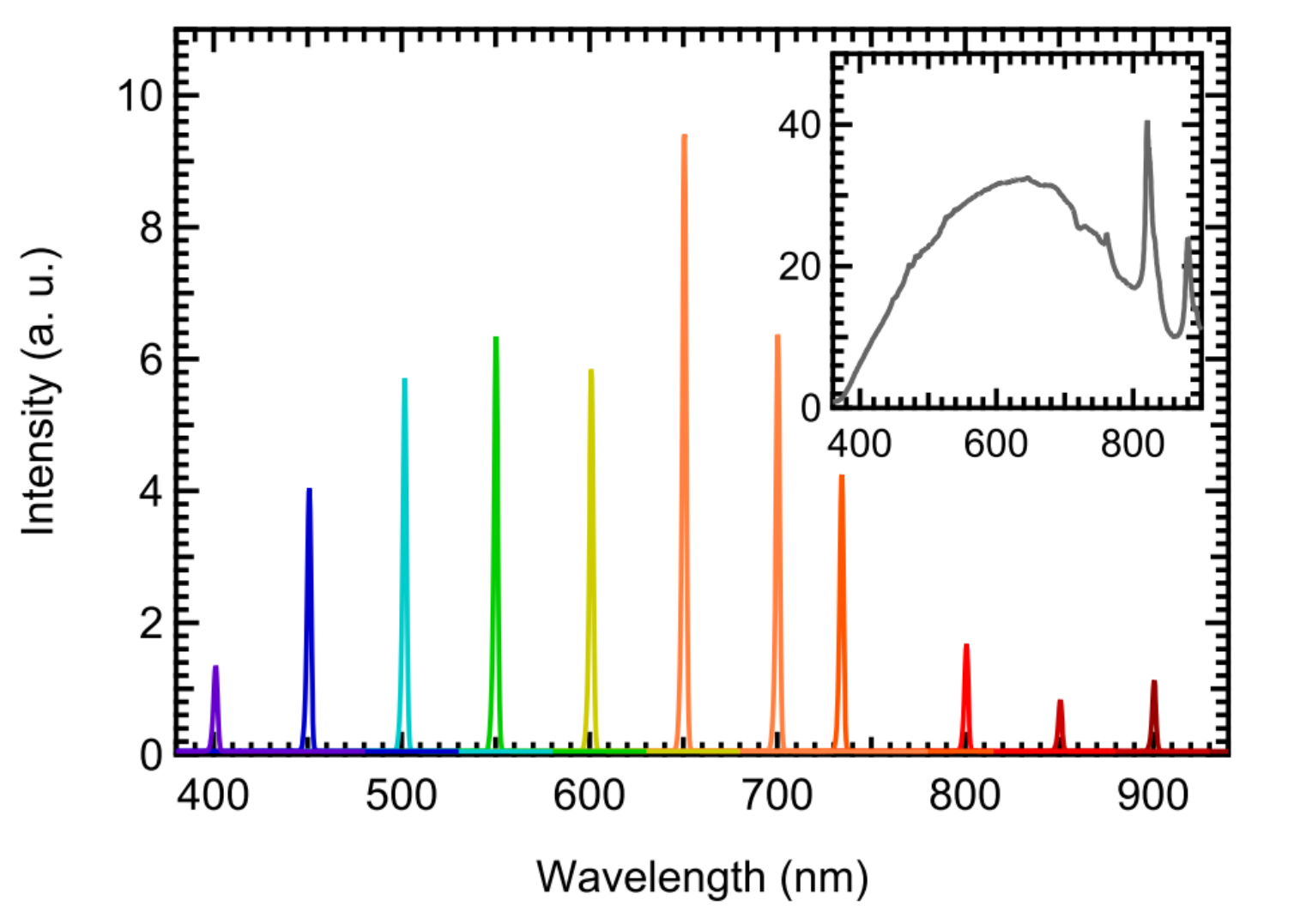}
\caption{Selected single-wavelength obtained after the light source  exits the monochromator. Insert: Broadband light source spectrum used for this experiment.} 
\label{fig:EXP2}
\end{figure}

For the tunable monochromatic magneto-transmission measurement, a combination of laser-driven broadband white lamp (EQ-99X from Energetiq Technology Inc.) and a mini monochromator (Manual Mini-Chrom Monochromators from Edmund Optics Inc.) was served as the light source, which is highly compact, with a total length of less than 30\,cm. The white light from the laser-driven lamp was guided by a 400\,$\upmu$m multi-mode fiber to the input slit of the monochromator. The white light was then dispersed by the 1200\,g/mm grating and sent to the output slit. By controlling the width of the output slit, it is possible to obtain monochromatic light with a full width half maximal of $\sim$2.5\,nm and output power of $\sim$30\,$\upmu$w.
The tunability of the light source wavelength is demonstrated in Fig.~\ref{fig:EXP2}, where several selected monochromatic light spectra are presented. The broadband laser-driven lamp spectrum  in the range of 400-900\,nm is shown as the insert of Fig.~\ref{fig:EXP2}. As for the detection part, a 100\,kHz avalanche photodetector (APD440A from Thorlabs Inc.) was used to measure the output power of the transmitted light from the sample. This photodetector has a high sensitivity and able to measure the light power down to 1.1\,pW with a saturation power of 1.54\,nW. In the Faraday rotation measurement, two linear polarizers were positioned above and below the sample to define and analyze the polarization of the light, thereby enabling the measurement of field-induced polarization rotation.

For the spectrally resolved magneto-transmission measurements, the optical path remained identical to that described above. The key difference was the use of a tungsten-halogen lamp (SLS201L/M from Thorlabs Inc.) as the light source, while a spectrometer (IsoPlane SCT320 coupled to an EMCCD from Princeton Instrument Inc.) was employed as the detector. The typical exposure time was 0.5\,ms. The average variation of the magnetic field during the exposure time of the spectrum is less than 2\% of maximal field, in which case the spectrum can be regarded as taken in an essentially constant magnetic field condition. The typical read-out time between two spectra is around 0.7\,ms, as shown in Supplementary Information\,Fig.\,\ref{SI:Fig:Readout}(b).

\section{\label{sec:level3}Result and discussion}

\subsection{\label{subsec:level32} Faraday rotation 
measurement }

\begin{figure*}
\includegraphics[scale=0.42]{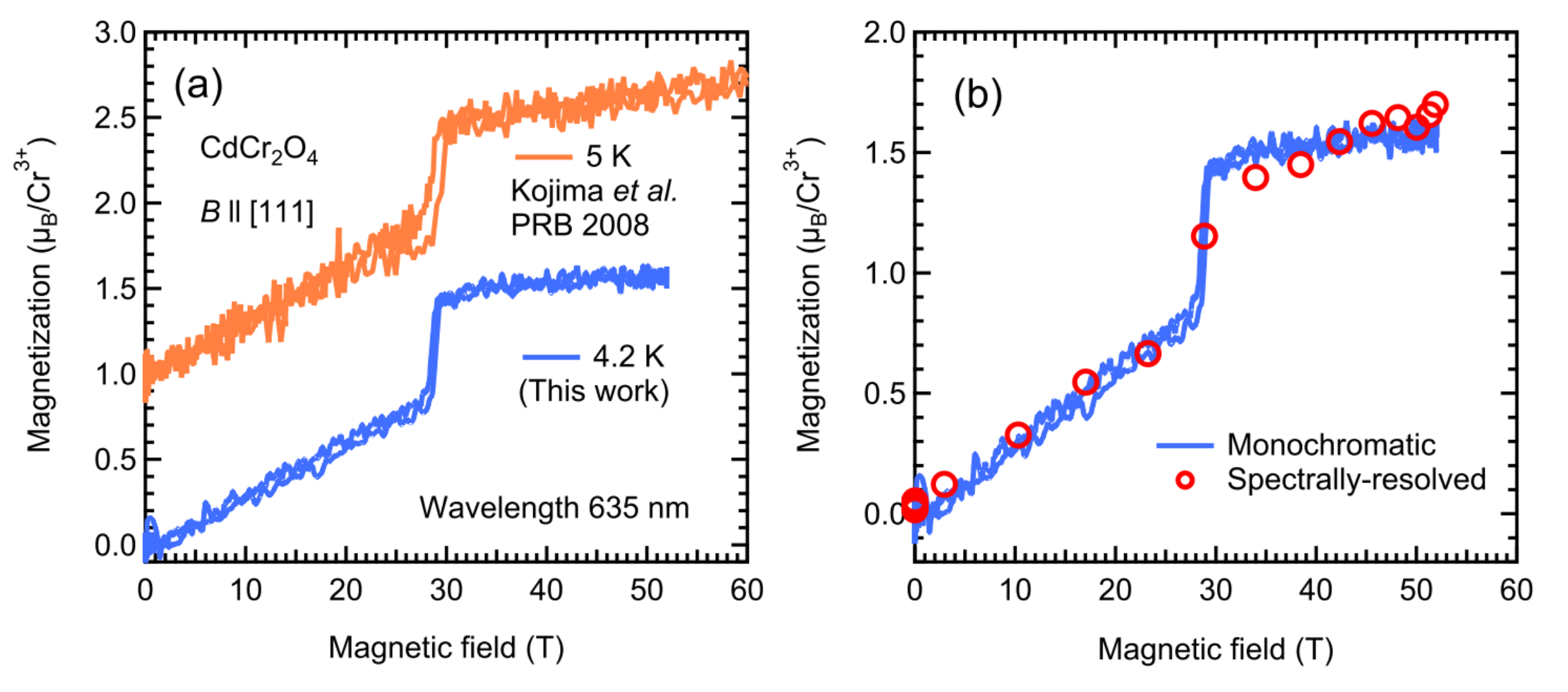}
\caption{\label{fig:ResultFaraday} (a) Normalized magnetization curve obtained at 4.2\,K compared with literature\cite{PhysRevB.77.212408}, the curves are vertically shifted for clarity. (b) Normalized magnetization using  single-wavelength or spectro-resolved magneto-transmission. 
}

\end{figure*}

Faraday rotation measurements were conducted on a 2\,mm diameter $\rm CdC{r_2}O_4$ sample at 4.2\,K with a magnetic field up to 52\,T ($B \parallel [111]$). To compare with previous Faraday rotation results on the same sample\,\cite{PhysRevB.77.212408}, we center the light source around  $\lambda$ = 635\,nm. After subtracting a linear background owing to diamagnetic contribution to the Faraday rotation angle from the quartz substrate, polarizer, and sample (See Supplementary Information Sec.\,\ref{SI:Sec:FaradayRotation}), the normalized magnetization is obtained and shown in Fig.\,\ref{fig:ResultFaraday}(a). The magnetization measured in this work increases linearly with the magnetic field up to 28\,T, where a sharp transition is then observed. After the transition, the magnetization is almost constant from 30 to 52\,T. This result is in excellent agreement with the previous report\,\cite{PhysRevB.77.212408}, as the two curves are almost indistinguishable from each other, highlighting the precision of this setup. 

The conventional method for determining the Faraday rotation angle $\theta$ relies on analyzing the field-dependent intensity of the $s$- and $p$-polarized light components, which are separated by a beam splitter and resolved by two orthogonal linear polarizers\,\cite{BookMag}. However, such an optical configuration cannot be implemented in our highly compact optical probe owing to limited space (see Fig.\,\ref{fig:EXP}(a)). To overcome this limitation, we developed an alternative method for measuring the Faraday rotation using our setup. Specifically, we perform two measurements under reversed experimental conditions, such that the Faraday rotation angle appears with opposite sign, while other experimental factors remain unchanged. By subtracting the two measurements, we isolate and extract the pure Faraday rotation signal. Since the sign of $\theta$ depends solely on the direction of the magnetic field $\mathbf{B}$ and/or the light propagation vector $\mathbf{k}$\,\cite{BookMag}, our method allows flexibility in reversing either of these to realize differential measurements, thereby enhancing the versatility of the system. A detailed description of this method is provided in Supplementary Information Sec.,\ref{SI:Sec:FaradayRotation}.

To highlight the advantages of our experimental setup, we compare the results with those obtained from spectrally resolved magneto-transmission measurements performed at $T$ = 1.5\,K. For each spectrum, we extract the intensity within a 2.5\,nm bandwidth centered at $\lambda$ = 635\,nm and plot it as a function of magnetic field. Although both curves exhibit a similar linear trend at low field, the sharp transition near 28\,T is not clearly resolved in the spectrally resolved data owing to the limited field resolution. Moreover, the high optical power of the white light source used in the spectrally resolved setup can induce significant sample heating, whereas the monochromatic configuration minimizes the heating effect owing to its much lower power. These observations demonstrate that tunable monochromatic magneto-transmission measurements are better suited for pulsed magnetic field experiments, particularly when high field resolution and low sample heating are critical.

\subsection{\label{subsec:level31} Tunable monochromatic magneto-transmission measurement }

\begin{figure*}
\includegraphics[scale=0.42]{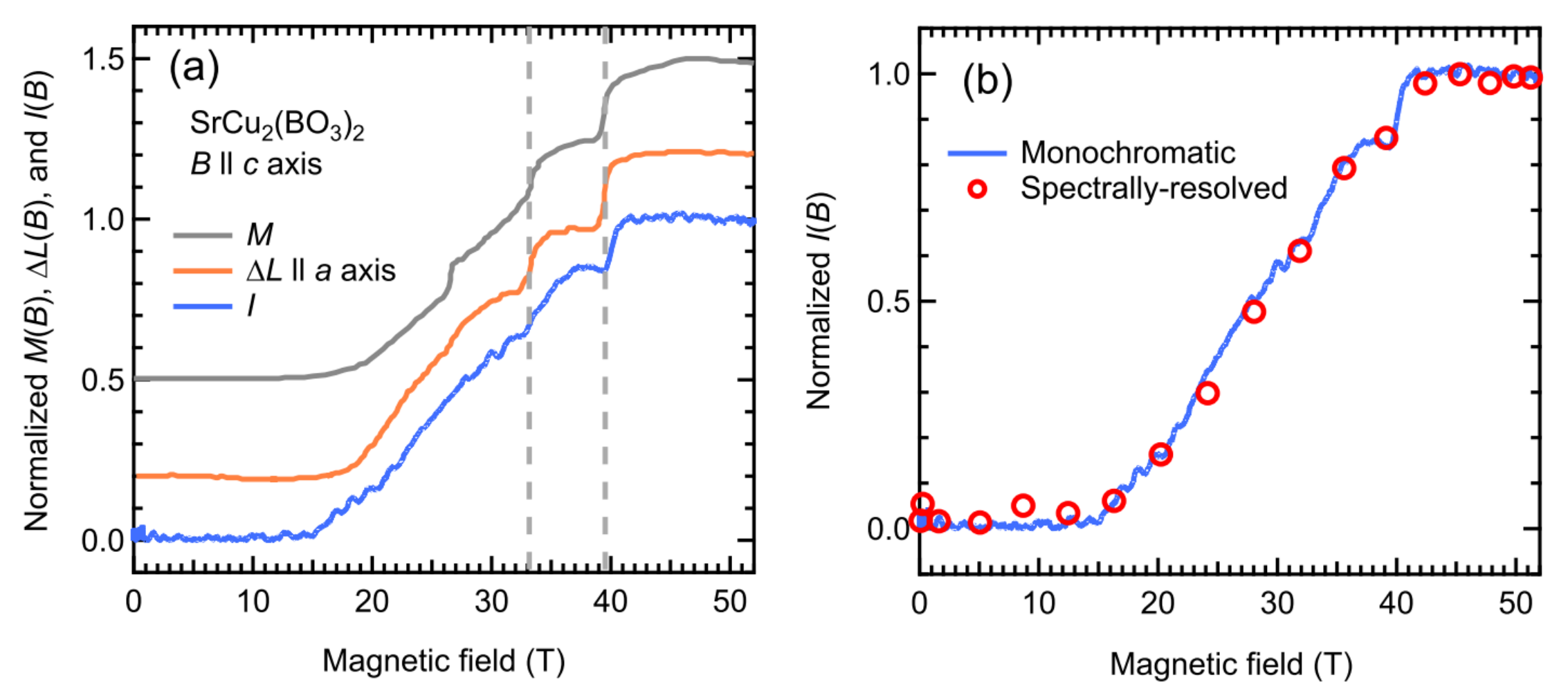}
\caption{\label{fig:Result1}(a) Normalised transmission signal as a function of field compared with the reported magnetization and magnetostriction data\cite{onizuka20001,radtke2015magnetic}, the curves are vertically shifted for clarity. (b) Normalized transmission signal from tunable monochromatic and spectrally-resolved magneto-transmission.
}
\end{figure*}

It is known that the band edge shift of SCBO in the magnetic field is strongly influenced by the short-range spin correlation\,\cite{PhysRevB.90.014405}. Furthermore, SCBO exhibits numerous phase transitions characterized by magnetization plateaus \cite{Kodama_2002} making this sample suitable to test the tunable monochromatic magneto-transmission technique.
The monochromatic light with center wavelength of $\lambda$ = 735\,nm was chosen to be close to the band edge to ensure the sensitivity to the band edge shift (see Supplementary Information Fig.\,\ref{SI:Fig:Readout}). The normalized transmitted light intensity of SCBO sample as a function of magnetic field up to 52\,T ($B \parallel c$ axis) with $T$ = 1.5\,K is shown in Fig.~\ref{fig:Result1} (a).

 The light intensity stays flat up to 17\,T and then increases linearly until 30\,T. This indicates the closing of the spin gap in SCBO and the non-magnetic singlet ground state becomes magnetic owing to the presence of triplets\cite{PhysRevB.90.014405}. At higher fields, kinks start to appear and a  plateau is well observed starting from 40\,T and staying flat until 52\,T. For comparison, we show in Fig.~\ref{fig:Result1} (a) the reported magnetization data\,\cite{onizuka20001} and transverse magnetostriction data ($\Delta L \parallel a$ axis)\,\cite{radtke2015magnetic}. All the features in magnetization and magnetostriction of SCBO are well captured by our magneto-transmission data, highlighting the sensitivity of the setup.\\

To precisely estimate the phase transitions of SCBO from the magneto-transmission data, we take a derivative of  this curve with respect to magnetic field, as shown in the Supplementary Information Fig.\,\ref{SI:Fig.SCBO_Magnetization}. The upper and lower panels of Fig.\,\ref{SI:Fig.SCBO_Magnetization} represent two sets of data obtained from the same measurement condition to ensure the reproducibility of the experiment. The magnetic field values of the peak positions are summarized in Table\,\ref{tab:result1} to compare with the critical fields of the sequence of phase transitions in SCBO from Onizuka et al.\,\cite{onizuka20001}. Comparing the critical field values in Tab.~\ref{tab:result1}, it is clear that the data from transmission and magnetization measurements exhibit excellent agreement with an error bar of 1\,T. This demonstrates the accuracy of our experimental setup.

As mentioned above, the tunable monochromatic magneto-transmission results successfully detected the sequence of phase transition of SCBO, and are well consistent with the previously reported magnetization and magnetostriction data. This high resolution and sensitivity of the technique is a big advantage when compared to the spectro-resolved magneto-transmission.  To demonstrate this, we plot the data points extracted from spectro-resolved magneto-transmission spectra at 735\,nm in Fig.\,\ref{fig:Result1}(b) as red dots (see Supplementary Information Sec.\,\ref{SI:Sec:spectro-resolved} for the spectrally-resolved data). owing to the read-out time of the CCD camera, the data points from the spectra are much less than those from the photodetector. The density of the experimental data points from the 100\,kHz bandwidth photodetector has 100 points per millisecond, whereas the spectrometer data has only 0.8 data points per millisecond. The fewer data points from the spectrally-resolved technique inevitably induce certain information loss for the measured system; for example, the magnetization plateau of SCBO between 39\,T and 42\,T is almost invisible for the spectro-resolved data. \\
 
The newly developed magneto-optical setup enables the precise detection of a sharp phase transition near 28 T in $\rm CdCr_2O_4$, as well as a sequence of magnetization plateaus in SrCu$\rm_2$(BO$\rm_3$)$\rm_2$ using pulsed magnetic fields with a pulse duration of only 36\,ms. 
In addition to its high-speed data acquisition and user-selectable wavelength tunability, the system is highly cost-effective, stable, and compact. Recent development of compact and portable pulsed magnet systems capable of generating fields up to 40\,T\,\cite{ikeda2024concise} has made high magnetic field measurement accessible in diverse laboratory environments. Integrating such a system with our optical setup opens new opportunities for high-resolution, field-dependent studies of magneto-optical responses, serving as a powerful and versatile platform to study unconventional magneto-optical phenomena in quantum materials.

\begin{table}
\caption{ \label{tab:result1} The critical fields of the sequence of phase transitions in SCBO from magnetization results \cite{onizuka20001} and from our $ dI(B)/dB$ signal. The error bar is estimated to be $\pm$1\,T.}
\begin{tabular}{|ll|c|c|c|}
\hline 

\multicolumn{2}{|l|}{Critical Field}                                            & $H_{C_1}$ (T)   & $H_{C_2}$ (T)   & $H_{C_3}$ (T)   \\ \hline

\multicolumn{1}{|l|}{Magnetization} & & 27 & 34 & 39 \\ \hline

\multicolumn{1}{|l|}{\multirow{2}{*}{\begin{tabular}[c]{@{}l@{}}Magneto- transmission\end{tabular}}} & Pulse 1 & 24 & 34 & 40 \\ \cline{2-5} 

\multicolumn{1}{|l|}{} & Pulse 2 & 23 & 34 & 40 \\ \hline
\end{tabular}
\end{table}

\section{Conclusion}
We have presented a newly developed magneto-optical setup for tunable monochromatic magneto-transmission measurements in non-destructive pulsed magnetic fields up to 52\,T. This system enables high-density transmission data acquisition as a function of magnetic field at user-selected wavelengths, while remaining highly cost-effective, stable, and compact in size. The performance of the setup was demonstrated by Faraday rotation measurements on a geometrically frustrated spin system $\rm CdCr_2O_4$ which enabled the detection of a sharp phase transition near 28\,T. Additionally, we investigated the band-edge behavior of the Shastry-Sutherland lattice antiferromagnet SrCu$\rm_2$(BO$\rm_3$)$\rm_2$, where the magneto-transmission and its derivative $dI(B)/dB$ exhibited a sequence of magnetic phase transitions consistent with previous magnetization and magnetostriction studies. These results validate the reliability and versatility of the system and confirm that our setup provides a robust and space-efficient solution for wavelength-tunable magneto-optical experiments under pulsed magnetic fields. 
This capability represents a key technological advancement for probing the precise magnetic-field dependence of magneto-optical phenomena and plays a vital role in the investigation of exotic magneto-optical effects in quantum materials.

\bibliography{Reference.bib}

\end{document}